\documentclass{elsart}
\usepackage{doublespace}
\def\be{\begin{equation}}
\def\ee{\end{equation}}
\def\ba{\begin{eqnarray}}
\def\ea{\end{eqnarray}}
\def\>{\rangle}
\def\<{\langle}
\def\n{\nonumber}
\def\lb{\left[}
\def\rb{\right]}
\def\<{\langle}
\def\>{\rangle}
\begin{document}
\begin{frontmatter}
\title{Uncertainty rescued: 
Bohr's complementarity for composite systems}
\author{Ilki~Kim and G\"{u}nter~Mahler}
\address{Institut f\"{u}r Theoretische Physik I, Universit\"{a}t Stuttgart\\
Pfaffenwaldring 57, 70550 Stuttgart, Germany\\
phone: ++49-(0)711-685-5100, FAX: ++49-(0)711-685-4909\\
email: ikim@theo.physik.uni-stuttgart.de}
\begin{abstract}
Generalized uncertainty relations may depend not only on the commutator 
relation of two observables considered, but also on mutual correlations, 
in particular, on entanglement. The equivalence between the uncertainty 
relation and Bohr's complementarity thus holds in a much broader sense 
than anticipated. 
\end{abstract}
\end{frontmatter}
%
\section{Introduction}
Heisenberg's uncertainty relation \cite{HEI27,ROB29} and Bohr's principle of 
complementarity \cite{BOH28} are at the heart of quantum theory. 
The relation of these two principles has, however, created much debate: 
the former is often interpreted to imply that one cannot detect 
for a given quantum state two conjugate observables with unlimited 
precision. The latter may be understood to mean \cite{SCU91} that 
a state with minimum dispersion of one observable (i.e. preparation of a 
respective eigenstate) implies maximum dispersion of the other (i.e. any of 
its eigenstates will be found with equal probability). Two conjugate 
variables like position $\hat{x}$\, and momentum $\hat{p}_x$\, are, at the 
same time, constrained by a typical uncertainty relation and 
complementarity in the above sense. Rather than considering these 
constraints as inherent properties of the quantum state, reference is 
often made to measurement disturbances (``random kicks''), as a 
quasi-classical explanation. This notion has a long history: 
In the famous Bohr-Einstein debates in 1920's \cite{BOH49}, 
Bohr invariably refuted Einstein's {\em{gedanken}} experiments 
(e.g. the `recoiling slit') by using the standard 
uncertainty relation to prove his complementarity. Thereafter similar 
conclusions have been postulated based on the momentum uncertainty 
in Heisenberg's microscope \cite{HEI30} and Feynman's electron-light 
scattering scheme \cite{FEY65}, respectively.\\
\hspace*{1ex}
In the 1990's this topic was revived by a {\em welcher-Weg} (`which-path') 
measurement scheme in quantum optics, theoretically proposed 
by Scully, Englert, and Walther \cite{SCU91}, and experimentally 
realized recently by Rempe and his collaborators \cite{DUE98} based on atomic 
beams and by Kwiat {\em et al.} \cite{KWI99} based on photon beams. 
These investigators claimed that the 
complementarity can be enforced without any uncertainty relation at work 
by exploiting quantum entanglement \cite{SCH35} 
between an atom (a photon) and a `which-path' 
detector: this detector allows to record the 
path by using the atom's different internal states (the photon's different 
polarisation states, respectively). 
The complementarity would thus appear as a more general and more fundamental 
principle than the uncertainty rule 
\cite{KNI98,ENG99}. However, this conclusion, `complementarity 
without uncertainty relation', has been questioned 
by Steuernagel \cite{STE99}. Based on a formal 
spin-$1/2$\, 
system, where two `paths' can be assigned to the two basis\linebreak
states, he showed that uncertainty and complementarity would mutually imply 
each other. Unfortunately he was unable to explicitly connect his results 
with the original proposal \cite{SCU91} and Rempe's scheme, respectively. 
It was already in 1994 when Storey {\em{et al.}} \cite{STO94,STO95} 
argued that the wave-particle 
duality should always be enforced by some momentum transfer as might be 
expected from the standard uncertainty relation. 
We intend to clarify this controversy 
by referring to generalized uncertainty relations for composite systems 
defined in a finite-dimensional state space.
%
\section{Generalized uncertainty relations}
Let us first state the general 
form of uncertainty relations: for a given state operator $\hat{\rho}$ and 
two observables $\hat{A}_j\,,\, j = 1, 2$\,, with $\< \hat{A}_j \> = 
\mbox{Tr} \left( \hat{\rho} \hat{A}_j \right)$\, denoting the respective 
expectation values, we define 
$\delta \hat{A}_j = \hat{A}_j - \< \hat{A}_j \> \hat{1}$\,, where $\hat{1}$ is 
the unit operator. Then, with the commutator
\be
\lb \delta \hat{A}_1\, ,\, \delta \hat{A}_2 \rb_{-}\; =\; 
\lb \hat{A}_1\, ,\, \hat{A}_2 \rb_{-}\; =\; 2 i\,\hat{B}\,,
\ee
the generalized uncertainty relations \cite{SCH30} read
\be
\label{uncertainty}
\left( \Delta A_1 \right)^{2} \left( \Delta A_2 \right)^{2}\; \geq\; 
\left| \left\< \hat{B} \right\> \right|^2 + \chi_{A_{1}A_2}^2\,,
\ee
i.e. the product of the variances $\left( \Delta A_j \right)^{2} = 
\left\< \left( \delta \hat{A}_j \right)^2 \right\>$\, is bounded from below 
by the sum of two terms: the expectation value of $\hat{B}$ and 
the symmetrized covariance
\be
\label{cov}
\chi_{A_{1}A_2}\, =\, {\displaystyle \frac{1}{2} \left\< \delta \hat{A}_1\, 
\delta \hat{A}_2 + \delta \hat{A}_2\, \delta \hat{A}_1 \right\>}\,.
\ee
The first term on the right hand side in (\ref{uncertainty}) restricts the 
(ensemble-) measurement outcomes with respect to two non-commuting observables 
$\left(\, \mbox{for which}\; \hat{B}\right.$\linebreak
$\left.\ne 0\, \right)$\,. The second term 
accounts for the influence of correlations: it has a classical 
analog \cite{SCH30} and may be unequal zero even for two 
commuting observables. 
For the two canonically conjugate operators $\hat{A}_1 = \hat{x}$ and 
$\hat{A}_2 = \hat{p}_x$\,, the commutator $\hat{B}$\, is proportional to 
the unit operator $\hat{1}$\,, so that \cite{SHA80}
\be
\label{unschaerfe}
\left( \Delta x \right)^{2} \left( \Delta p_x \right)^{2}\; \geq\; 
\frac{1}{4} \hbar^2 + \frac{1}{4} \left\<\psi\left| \lb \delta \hat{x}, 
\delta \hat{p}_x \rb_{+} \right|\psi\right\>^2\,.
\ee
The inequality refers to one given initial state, not to a sequential or 
even `simultaneous' measurement of $\hat{x}$ and $\hat{p}_x$ on the same 
individual system. 
The last term in equation~(\ref{unschaerfe}) is usually discarded as an 
explicitly state-dependent correction. Note, however, that the first 
(state-independent) term is not generic: such a term cannot occur in any 
finite-dimensional Hilbert-space as the commutator must be traceless. 
But explicit state-dependence renders measure-\linebreak
ment-induced random kicks 
as the origin of quantum uncertainty a much less convincing concept.
%
\section{Which-path detection model}
It has been stated that at least the `historical' which-path detectors in a 
double-slit configuration could be `explained' in terms of the canonical 
relations (\ref{unschaerfe}). We doubt that: in Feynman's light-scattering 
arrangement \cite{FEY65} a light beam is supposed to interact with the 
electrons after they have passed through the double-slits to reveal their 
paths. Though the momentum `kicks' by the photons might explain why the 
interference patterns on the screen are washed out \cite{STO94,STO95}, but it 
remains open why this continuous random perturbation $\Delta p_x$ should 
produce exactly 2(!) alternative patterns (the sub-ensembles originating from 
slit 1 as if slit 2 were closed and from slit 2 as if slit 1 were closed). 
The photon momentum appears to play a role, though, when Feynman proposes 
to reduce the perturbation by reducing the photon momentum, thus increasing 
the wavelength. The `which-path' detector ceases to work when the slits 
(the two paths) can no longer be resolved, and the interference patterns 
reappear. However, rather than being a result of reduced `random kicks', this 
should be seen as a logical consequence of the fact that in this limit the 
`which-path' detector is unable to distinguish the two paths, so that the 
necessary correlation cannot build up (see below). 
Similar arguments should apply to Einstein's recoiling slit 
\cite{BOH49}, as well as to early experimental realizations \cite{EIC93}.\\
\hspace*{1ex}
Interference may result when one final state of a quantum object can be 
reached in at least two different ways, e.g. from two different initial 
states. Typical experimental settings involve pre-selected paths. 
Even though spatial- or momentum-coordinates form a continuous set, 
the experimental design for a `which-path' detection typically reduces 
that space to discrete (usually\linebreak
two) alternatives, just as the electronic 
levels of an atom, say, can be selected to simulate a two-level system 
(pseudo-spin) \cite{ENG96}. 
Such an effective two-level system is most conveniently described in terms 
of the Pauli operators $\hat{\sigma}_j\,,\, j = x, y, z$\, with 
$\left(\hat{\sigma}_j\right)^2 = \hat{1}$\, and e.g. 
$\lb \hat{\sigma}_x , \hat{\sigma}_y \rb_{-} = 2 i\,\hat{\sigma}_z$\,. 
As a consequence the state $\hat{\rho}$ can be specified by the Bloch vector 
$\sigma_j = \left\< \hat{\sigma}_j \right\>$ and its variance by 
$0 \leq \left( \Delta \sigma_j \right)^2 = 1 - \left( \sigma_j \right)^2 
\leq 1$\,. Finally, using\, $\chi_{\sigma_x \sigma_y} = 
- \sigma_x\, \sigma_y$\,, equation~(\ref{uncertainty}) now implies
\be
\left( \Delta \sigma_x \right)^2 \left( \Delta \sigma_y \right)^2\; \geq\; 
\left( \sigma_z \right)^2 + \left( \sigma_x \right)^2\, 
\left( \sigma_y \right)^2\,.
\ee
Further inequalities are obtained by permuting the indices $x, y, z$\,. This 
inequality constitutes a relation between first and second moments; there is 
no state-independent term. Nevertheless, we see that perfect knowledge of 
$\hat{\sigma}_z$\,, say, $\left(\sigma_z = \pm 1\right)$\, leads necessarily 
to complete ignorance about $\hat{\sigma}_x$\, and $\hat{\sigma}_y$\, 
$\left(\Delta \sigma_x = \Delta \sigma_y = 1,\, \mbox{i.e.}\; 
\sigma_x = \sigma_y = 0
\right)$\,. Obviously, these three observables are pairwise complementary, 
without being conjugate like the canonical variables in (\ref{unschaerfe}).\\
\hspace*{1ex}
In the following we will extend these considerations to composite systems 
(composed of distinguishable subsystems). While the local observables 
referring to different subsystems are always commutative, they both turn out 
to be complementary to quantum correlation (the covariance term 
in (\ref{uncertainty})). This conclusion can be reached in various ways; here 
we will show that it can be derived from an appropriate uncertainty 
relation.\\
\hspace*{1ex}
Interference between a double-slit results from the superposition of two 
different paths (modes). The two paths will be identified as the eigenstates 
of $\hat{\sigma}_z$\,, 
$\hat{\sigma}_z\, |\pm z\> =\, \pm |\pm z\>$\,. 
Interference then requires the preparation of superposition states 
like\, $A |+z\> + B |-z\>$\,. In the concrete double-slit 
experiment,\, $A, B$ will depend on the spatial coordinates on the screen. 
Here we restrict ourselves to the form
\be
\label{coherent}
|\varphi\>\; =\; {\displaystyle \frac{1}{\sqrt{2}}\, 
\left(\, |+z\> + e^{i \varphi} |-z\>\, \right)}\,.
\ee
The states 
$|\varphi=0\> = |+x\>$\,, $|\varphi=\pi\> = |-x\>$\, are formal eigenstates of 
$\hat{\sigma}_x$\,. At any point of the screen we measure 
$|\varphi=0\>$\,, i.e. the corresponding probability $P_0(\varphi) = 
\left| \< 0 | \varphi \> \right|^2 = 
\frac{1}{2}\, (1 + \sigma_x) = \frac{1}{2}\, (1 + \cos \varphi)$\,, which 
constitutes our idealized interference fringes with fringe visibility
\be
\mathcal{V}\; =\; 
\frac{{\vspace*{\fill}P_0(\varphi)}_{\mbox{{\footnotesize{max}}}}\, -\, 
{\vspace*{\fill}P_0(\varphi)}_{\mbox{{\footnotesize{min}}}}}
{{\vspace*{\fill}P_0(\varphi)}_{\mbox{{\footnotesize{max}}}}\, +\, 
{\vspace*{\fill}P_0(\varphi)}_{\mbox{{\footnotesize{min}}}}}\; =\; 1\,.
\ee
The quantitative measure, $\mathcal{D} = 0$\,, for the 
$\left|\pm z\right\>$-distinguishability is then obtained 
from $\mathcal{D}^2 + \mathcal{V}^2 = 1$\, \cite{ENG96}.\\
\hspace*{1ex}
While the interference pattern becomes visible only as an ensemble result, 
it is, nevertheless, a single-particle property. Ideal `which-path' 
detection therefore requires a single-particle sensitivity. 
By such a `which-path' detection scheme, a physical label is 
introduced to mark those `paths'\, i.e. to make them `distinguishable'. 
For this purpose 
we introduce a second two-level system (subsystem 2\,, e.g. an 
internal atomic two-level system or the polarisation\linebreak
states of a photon), 
as proposed in \cite{SCU91} (see also \cite{REI99}): 
the `which-path' detection requires to build up a strict 
correlation or anti-correlation between the paths, $|\pm z\>^{(1)}$\,, 
and the marker states, $|\pm z\>^{(2)}$\,, e.g. $|+z\>^{(1)} \otimes 
|+z\>^{(2)}\,,\, |-z\>^{(1)} \otimes |-z\>^{(2)}$\,. 
Because of the linearity of quantum mechanics the local coherent state of 
subsystem~1 evolves into
\be
\label{entangled}
|\psi(1,2)\>\, =\, \frac{1}{\sqrt{2}}\, 
{\displaystyle \left(\; |+\hspace*{-.1cm}z\; , +z\>\, +\, 
|-\hspace*{-.1cm}z\; , -z\>\; \right)}\,,
\ee
where the first index refers to subsystem 1 (path index), and the second one 
to subsystem~2 (marker). This state can formally be obtained by means of a 
quantum-controlled NOT operation \cite{MAH98}: after subsystem~2 has been 
prepared in a standard state, say, $|-z\>^{(2)}$\,, we apply the unitary 
transformation\linebreak
$|-z\>^{(2)} \to 
|+z\>^{(2)}$ if the state of subsystem 1 is $|+z\>^{(1)}$, no change otherwise 
(transition table: $|+z,+z\> \to |+z,-z\>\,,\, 
|+z,-z\> \to |+z,+z\>\,,\, |-z,+z\> \to |-z,+z\>\,,\, 
|-z,-z\> \to |-z,-z\>$\,).
%
\section{Uncertainty versus complementarity}
The total system can be described in terms of the two-particle operators, 
$\hat{K}_{jk} = \hat{\sigma}_j^{(1)}\, \otimes\, \hat{\sigma}_k^{(2)}\,,\; 
j, k = x,  y, z$\,, the single-particle operators, 
$\hat{K}_{j0} = \hat{\sigma}_j^{(1)}\, \otimes\, \hat{1}^{(2)}$\,, 
$\hat{K}_{0k} = \hat{1}^{(1)}\, \otimes\, \hat{\sigma}_k^{(2)}$\,, and 
$\hat{K}_{00} = \hat{1} = \hat{1}^{(1)}\, \otimes\, \hat{1}^{(2)}$ 
\cite{MAH98}. The incompatibility between the single- and the two-particle 
operators can be specified by
\ba
\lb \hat{K}_{jj}\,, \hat{K}_{0k} \rb_{-}\; &=&\; 
\hat{\sigma}_{j}^{(1)} \otimes 
\lb \hat{\sigma}_{j}^{(2)}\,, \hat{\sigma}_{k}^{(2)} \rb_{-}\, \ne\, 0 
\;\;\;\; \mbox{for}\;\;j \ne k \n\\
\lb \hat{K}_{jj}\,, \hat{K}_{l0} \rb_{-}\; &\ne&\; 0 \;\;\;\; 
\mbox{for}\;\;j \ne l\,.\n
\ea
The single-particle 
operators acting on different subsystems obviously commute. 
Their covariance, equation~(\ref{cov}), is \cite{GRA99}
\be
\chi_{\sigma_j^{(1)} \sigma_k^{(2)}}\; =\; 
K_{jk} - \sigma_j^{(1)} \sigma_k^{(2)}\; =\; M_{jk}\,,
\ee
where $K_{jk} = \left\< \hat{K}_{jk} \right\>$ describes two-particle 
correlations. One easily convinces oneself that in general 
$0 \leq |M_{jk}| \leq 1$\,, while\, $M_{jk} = 0$\, for product states 
\cite{SCH95}. Accordingly, our new inter-subsystem uncertainty 
relations are given by
\be
\label{uncertain}
\left( \Delta \sigma_j^{(1)} \right)^2\, \left( \Delta \sigma_k^{(2)} 
\right)^2\; \geq\; \left| M_{jk} \right|^2\,,\;\;\; j, k = x, y, z\,.
\ee
In this case the uncertainty relations in the conventional form, i.e. 
discarding the covariance term, would be absolutely insufficient. 
This inequality is used here to assess quantum mechanical properties 
even though its form would apply also to any pair of classical statistical 
variables $\sigma_j^{(1)}\,, \sigma_k^{(2)}$\,. For pure\linebreak
states, 
$M_{jk}$ implies non-local quantum correlations, and these quantum effects 
can still be manifest in equation (\ref{uncertain}): for the case 
$\left|M_{jk}\right| = 1$\,, which means strict (anti)-correlation, one 
obtains a maximum ignorance of $\hat{\sigma}_j^{(1)}$\, and 
$\hat{\sigma}_k^{(2)}$\, $\left(\Delta \sigma_j^{(1)} = 
\Delta \sigma_k^{(2)} = 0\right)$\,. On the other hand, the perfect knowledge 
about $\hat{\sigma}_j^{(1)}$\, and $\hat{\sigma}_k^{(2)}$\, leads to maximum 
uncertainty in the correlation, $\left|M_{jk}\right| = 0$\,.\\
\hspace*{1ex}
The state (\ref{entangled}) (with distinguishability 
$\mathcal{D} = 1$\, for the `which-path' information) is a joint 
eigenfunction of the commuting set $\hat{K}_{jj}\,,\, j = x, y, z$\, and has 
$\left| M_{ij} \right| = \delta_{ij}$\,. It thus shows zero fringe visibility, 
$\mathcal{V} = 0$\, $\left(P_0(\varphi) = \mbox{const.}\right)$\,. 
The preparation of the entangled state 
implies that each subsystem will be in a non-pure state. This fact may seem 
to justify an interpretation in terms of a `mixture' resulting from some 
randomization (e.g. due to `photon kicks', 
see \cite{FEY65,STO94,STO95,HOW95}). But 
this picture becomes inconsistent as we are able to completely remove the 
alleged randomization by post-selection (`quantum erasure'). 
To see this we observe that equation (\ref{entangled}) can be rewritten as
\ba
|\psi(1,2)\> &=& \frac{1}{\sqrt{2}}\, 
{\displaystyle \left(\, |+\hspace*{-0.8mm}x\; , +x \>\, +\, 
|-\hspace*{-0.8mm}x\; , -x \>\, \right)}\,.\n\vspace*{-0.1cm}
\ea
If we now sort out $|+\hspace*{-0.5mm}x\>^{(2)}$\, or 
$|-\hspace*{-0.5mm}x\>^{(2)}$\, - and only then - we recover the 
respective interference patterns, as proposed by Scully {\em et al.} 
\cite{SCU91}. This result is in accord with the 
uncertainty relation (\ref{uncertain}), as this sorting out requires a 
measurement and after having obtained 
$\sigma_x^{(1)} = \sigma_x^{(2)} = +1$\, or $-1$\,, one of which is 
post-selected, we obtain 
$\Delta \sigma_x^{(1)} = \Delta \sigma_x^{(2)} = 0$ and thus $M_{xx} = 0$\,, 
indicating a product state with local coherence 
(cf. equation (\ref{coherent})).\\
\hspace*{1ex}
Other marker states would also work. As an example we consider a two-spin 
system (subsystem 2,3) and choose as the initial state
\ba
&&|\psi(2,3)\> = \frac{1}{\sqrt{2}}\, {\displaystyle \left(\, 
|+\hspace*{-0.8mm}z\; , +z \>\, +\, |-\hspace*{-0.8mm}z\; , -z \>\, \right)}\,.
\n
\ea
If the coupling is the same as before (CNOT between subsystem 1 and 2), 
we have
\ba
\hspace*{-.9cm}&&|\psi(1,2,3)\> =\n\\
\hspace*{-.9cm}&&\frac{1}{2} \left( |+z\>^{(1)} \otimes 
\left( |-\hspace*{-0.8mm}z, +z\> + |+\hspace*{-0.8mm}z, -z\> \right) + 
|-z\>^{(1)} \otimes \left( |+\hspace*{-0.8mm}z, +z\> + 
|-\hspace*{-0.8mm}z, -z\> \right) \right)\,,\n
\ea
where the two Bell states in subspace (2,3) now play the role of an effective 
two-state subsystem. Note that the subsystems 2 and 3 are both in a `mixed' 
state for each alternative `path', carrying no local information! 
Such a situation has been investigated 
by Kwiat {\em et al} \cite{KWI99}.
%
\section{Summary and discussion}
We have tried to convince the reader that - as we include more general 
scenarios referring to composite systems - $(1)$\, complementarity and 
uncertainty relations can still strictly be related, $(2)$\, quasi-classical 
explanations for the origin of uncertainty (complementarity) in terms of 
measurement-induced ``random kicks'' become untenable. 
Recent `which-path' measurement schemes challenge the strict interrelation 
between uncertainty and complementarity: 
it is argued that entanglement may lead to 
complementarity without uncertainty relation. 
In fact, single-particle coherence $(M_{jk} = 0)$\, and two-particle 
coherence derived from entanglement $(M_{jk} \ne 0)$\, 
are complementary \cite{GRE93}.\\
\hspace*{1ex}
Here we noted that generalized uncertainty relations for composite 
quantum systems will, in general, depend not only on the commutator 
of the\linebreak
two observables considered, but also on mutual 
correlations, in particular, on entanglement. It thus follows that the 
equivalence between the set of uncertainty relations and 
Bohr's complementarity holds in a much broader sense than believed up to now. 
Equation~(\ref{uncertain})\, is the pertinent uncertainty relation 
underlying binary `which-path' measurements, not the intra-subsystem 
relation~(\ref{unschaerfe}). The former is but one out of a 
large class of hitherto unexplored inter-subsystem uncertainty relations, 
which can be generalized to any subsystem size and for 
which entanglement should turn out to be a natural ingredient. 
These inequalities between expectation values would be confirmed 
by appropriate ensemble measurements. But their interpretation in terms of 
`classical' measurement disturbances - already of limited qualitative value 
in simple cases - has to be abandoned in the general case:\\
\hspace*{1ex}
For a composite quantum system in a pure-state, 
entanglement implies that each subsystem is in a non-pure state, which would 
justify an interpretation in terms of a `mixture' 
resulting from some randomization. However, even though the maximum entangled 
state has lost its local interference properties, it exhibits, when correlated 
with the `which-path' measurement outcome, the same pattern as generated 
by either slit, without any additional randomness.\\
\hspace*{1ex}
Rather than talking of local `random kicks', 
we may say that each measurement `projects' the composite 
system in a correlated fashion into one of the states 
$|\pm z\>^{(1)}$ and thus subsystem 1 into one of the alternate paths. 
%
\section*{Acknowledgements}
We thank J.~Gemmer, Dr.~C.~Granzow, A.~Otte, P.~Pangritz, and F.~Tonner 
for fruitful discussions. 
One of us (I.~K.) acknowledges Dr.~O.~Steuernagel and Dr.~B.-G.~Englert 
for helpful discussions during the workshop `Entanglement and Decoherence' 
Gargnano, Italy, 1999.
\end{document}